
\input phyzzx
\pubnum={PUPT--1353}
\date={November 1992}
\titlepage
\title{{\bf A VIEW FROM THE ISLAND}\foot{Talk given at the 3rd International
Symposium on the History of Particle Physics, SLAC, June 24-27, 1992.}}

\author{A. Polyakov}
\address{Princeton University\break
Department of Physics\break
P.O. Box 708\break
Princeton, New Jersey 08544\break}

\abstract
This is a brief and subjective description of ideas which formed our
present field-theoretic understanding of fundamental physics.

\endpage

The ``island'' in the title of this article means two things--the Soviet
Union and my own mind.  Partial isolation from the larger physics community
had considerable effect on my work, both positive and negative.  While the
negative aspects of it are obvious, the good thing about isolation is that
it gives independence, reduces the danger of being swept by the
intellectual ``mass culture.''

The beginning of modern field theory in Russia I would associate with the
great work by Landau, Abrikosov and Khalatnikov [1].  They studied
the structure of the logarithmic divergencies in QED and
introduced the notion of the scale-dependent coupling.  This scale
dependence comes from the fact that the bare charge is screened by the
cloud of the virtual particles, and the larger this cloud is the stronger
screening we get.  They showed that at the scale r, the coupling has the
form
$$
\alpha(r)\infty \ \ 1/log \ r/a \ .
$$
where $a$ is the minimal cutoff scale.  Similar and, in some respect,
stronger results have been obtained by Gell-Mann and Low [2] who
discovered the ``renormalization group'' equation:
$$
{d\alpha\over d log\, r/a}\ =\ C_1 \alpha^2 + C_2 \alpha^3 + \cdots
$$

The catastrophic consequence of these results was that as $a \to 0$ (no
artificial cutoff) one obtains ``Moscow zero''--total vanishing of
interaction.  Immediately after that the search for different
renormalizable theories was started in an attempt to find antiscreening (or
as we would say today--asymptotic freedom).  The only finding at that time
had been the 4-fermion interaction in 2 dimensions (Anselm [3]).  This
caused well-known  pessimism towards field theory.  For the reasons
described below and also because I was seven years old in 1953, I have
never shared this pessimism.

Instead, I was very excited, when entering
physics in the early sixties, by this work and also by the marvelous ideas
of Nambu and Jona-Lasinion [4]  and Vaks and Larkin [5] who traced the
analogy between the 4 fermion masses and gaps in superconductors.  In the
USSR these works were considered as garbage, but they resonated with my strong
conviction, which I still hold today, that the really good ideas should
serve in many different parts of physics.  Even more than that--the
importance of the idea for me is measured by its universality.

As a result, starting from 1963, Sasha Migdal and I were involved in the
infinite discussions about the meaning of spontaneous symmetry breaking.
We had moral support from Sasha's father, a brilliant physicist and great
man, who has been almost the only one taking these ideas seriously.
At about the same time Larkin explained to us the physical origin of
massless particles (the ``Goldstone theorem'') and said that in the case of
long-ranged forces, as in superconductors, they don't occur, although exact
reasons for that have not been clear.

Sasha and I started to analyze
Yang-Mills theories with the dynamical symmetry breaking and in the spring
of 1965 came with the understanding that the massless particles must be
eaten by the vector mesons, which become massive after this meal.

We had many troubles with the referees and at seminars, but finally our
paper was published (Migdal--Polyakov [6]).  We did not know, until very
much later about the work on ``Higgs Mechanism'' which has been done in the
West at about the same time, or slightly earlier.

A little later I became
interested in the work on critical phenomena which was done by Pokrovsky
Patashinsky, Kadanoff and Vaks and Larkin.  It was quite obvious to me that
the critical phenomena are equivalent to relativistic quantum field theory,
continued to imaginary time. I felt that they provided an invaluable
opportunity to study elementary particle physics at small distances. The
``imaginary time'' didn't bother me at all; on the contrary I felt that it
is the most natural step, ultimately uniting space and time, and making the
ordinary time just a matter of perception.

With the use of the ingenious technique, developed by Gribov and Migdal
[7] in the problem of reggeons, I found connections between
phenomenological theory and ``bootstrap'' equations (Polyakov [8]).
 Sasha Migdal did very similar work independently.  There
was also something new--I formulated ``fusion rules'' for correlations,
which we  now would call operator product expansion [9].  I had mixed
feelings when I found out later that the same rules at the same time and in
more generality have been found by L. Kadanoff [10] and K. Wilson [11].

The paper by Wilson also overlapped with the project in which I was deeply
involved at the time.  It was an idea to describe elementary particles at
small distances, using renormalizable field theories.  I considered the
processes of the deep inelastic scattering and $e^+e^-$ annihilations, and
was able to prove that they must go in a cascade way, by forming few heavy
virtual objects, which I called ``jets'' and then by repeating the process
with lighter and lighter jets until we stop with real particles.  The
picture was inspired by Kolmogorov theory of turbulence.  I was able to
show that these processes are described by what is called now
``multifractal'' formulas and made predictions for the violations of
Bjorken scaling.  I considered both a scale-invariant (fixed-point) regime
with anomalous dimensions and a logarithmic regime which was easier to deal
with.

As a mathematical model I used $\lambda \phi^4$-theory, with the wrong sign
of $\lambda$.  However, looking through my old notes, I see that it was
just a toy model for me with no anticipation that the asymptotic freedom is
a real thing.  I thought at that time that anomalous dimensions are just
small numbers, like they are in the theory of phase transitions.  In any
case these papers [12] give a correct picture of the deep
inelastic processes in any renormalizable field theory, predicting the
pattern of the Bjorken scaling violation, the jet structure, the
multiplicity distribution (later called KNO-scaling).  In the beginning of
(1973) I finished the paper on the conformal bootstrap (Polyakov [13]),
but postponed its development for 10 years because I had heard in May 1973
about the results of Gross, Wilczek [14] and Politzer [15].  After a
short check it became clear to me that this is ``the'' theory.  All my old
statements about deep inelastic scattering were true in this case, but also
could be made much more concrete, since the coupling was small at short
distances.

It was not much of a challenge by then to elaborate this side of the
subject, and I turned to the large distance problem.  I was impressed by a
simple comment by Amati and Testa [16] that if you neglect $F^2_{\mu\nu}$
term in the gauge lagrangian, you obtain a constraint that the gauge
current is zero, the fact they associated with confinement.  In order to
make quantitative sense of this argument, I constructed a lattice version of
the gauge theory, in which the neglect of $F^2_{\mu\nu}$ is a well defined
approximation.  At the beginning of 1974 I gave a few talks on lattice
gauge theory, but never published it, since the preprint by K. Wilson came
in at this time.  It was clear that Ken had deeper understanding of the
subject of confinement, and I decided to do more work before publishing
something.

I kept thinking about the beautiful work by Berezinsky [17], in which he
showed very clearly how vortices and dislocations in two-dimensional
statistical mechanics create phase transitions. It was clear, that
confinement may be related to the fact that similar ``dislocations''
disorder the vacuum and create finite correlation length.  But what are
these ``dislocations'' in the gauge theory?

At this point I recalled my
conversation with Larkin at 1969 about Abrikosov vortex lines.  We
discussed whether they are normal elementary excitations appearing as
poles of the Green's functions. As it often happens, the discussion led
nowhere at that time but was helpful five years later.  What also helped
was my fascination with the work on solitons in the integrable systems,
being done by Zakharov and Faddeev at that time.  Actually Faddeev and
Takhtajan considered sine-gordon solitons as quantum particles.  What had
been far from clear was the extent to which these results were tied to the
specific models with complete integrability.

After brief but intense work in the spring of 1974, I arrive at two results
simultaneously. First, I found a nonabelian generalization of the Abrikosov
vortex in 3D and realized that it must be an elementary particle with
non-trivial topology.  A question, asked by L. Okun during my talk helped
me to realize that the topological charge is in fact magnetic charge.  The
same work was done simultaneously by G. 't Hooft.

While the possibility of
the magnetic poles has been envisaged by Dirac in the 30's, from our work
it follows that magnetic charges are inevitable in any reasonable unified
theory.  I am quite certain that they really exist.  How, when and if they
will be found is another matter.

The second result, published only a year later was that the same monopoles
play the role of the ``dislocations'' mentioned above in the $2+1$
dimensional gauge theories and indeed lead to confinement.  It took almost
a year to gain confidence in this result and to  find the dislocations in 4
dimensions.  In the abelian case, these dislocations turned to be just the
world lines of magnetic monopoles and I predicted the phase transition
leading to confinement (in $3+1$ case).
In the ($2+1$) case the confinement was the only phase (Polyakov [18]).
In the non-abelian ($3+1$) case it was necessary to find a novel solution
of the Yang-Mills equation in imaginary time and then to investigate its
influence on the vacuum disorder.  I suggested this problem to my
colleagues, Belavin, Schwartz and Tyupikin during some summer school and
together we have found the required solution.  Even before that, when I
discussed the problem with S. Novikov and asked him about the topology
involved in it, he told me about Chern classes.

I had never learned topology before and was somewhat scared by this subject.
I thought that my spatial imagination is not adequate for it.  At present,
I think that in topology just as in physics the more important quality
is the ``temporal'' imagination, also called ``intuition,'' the sense of
how things should be related in time.

Anyway, we had a solution (which obtained later the name ``instanton'') but
its effect on confinement turned out to be unclear because of strong
fluctuations of large instantons. That is why we don't have the theory of
confinement even today.  Nevertheless, instantons turned out to be
interesting beasts.  In the same summer  of 1975, Gribov noticed that they
can be interpreted as tunneling events between the different vacua.  It
became clear that the vacua in gauge theory were labeled by the integers
and the instanton was the process of jumping from one vacuum to another.

That was later rediscovered by other people.  Inspired by Gribov's remark,
I started to analyze the relation of the instantons to the axial anomaly,
when I heard about the beautiful results by 't Hooft, who had shown that
the tunneling, mentioned above, lead to the baryon number non-conservation
and to the solution of the $\eta$-mass problem.  I kept trying to solve the
confinement problem, playing with different physical settings.  In
particular I considered the temperature dependence of the gluon system,
and found a rather surprising (at that time) deconfining phase transition
(Polyakov [19]).

L. Susskind came to the same conclusion independently (L. Susskind [20]).
There are three interesting points about this work.  First, it has demonstrated
that the symmetry group, responsible for confinement is the center of the
gauge group, which breaks in the process of deconfinement.
Second, and more important, is that the natural description of the
deconfinement could be given in terms of condensing strings.  Third, I have
realized that temperature can alleviate tunneling and increase the
baryon number non-conservation via 't Hoof process (Polyakov [21]).
 The same idea occurred to L. Susskind.

The details, however, have been
worked out only in the 80's by many people.  I believe that at present this
is the most dramatic manifestation of the instanton structure of the
vacuum.

Since strings appeared so naturally in QCD, I turned to
string theory. First, I tried to use the equations in the loop space
(Polyakov [22]).  These loop equations still look interesting to me,
although very few results followed from them.  In particular, as was
noticed by A. Migdal, the equations simplify drastically in the large N
limit.  Migdal and Makeenko [23] showed how to reproduce perturbation
theory in this approach.  Unfortunately, we still don't know how to solve
these equations, but expect that the solution must be some kind of string
theory.

The fact that thirteen years of hard work didn't bring the
solution should not discourage us.  Problems in physics become more deep
and difficult and take more time than before.  For comparison, remember how
much time it took to solve some of the celebrated Hilbert mathematical
problems.  This is an inevitable consequence of the maturity of the
subject.

Incidentally, the work on instantons, which originated in complete
mathematical ignorance, seems to have influence on mathematics.  In the
hands of mathematical grand masters it helped to solve long standing
problems in topology of four-dimensional manifolds, and led to the link
between quantum field theory and topology.  That  shows that the notion of
``universality'' of good ideas should, perhaps include the realm of
mathematics.

We come (in the proper time of this article) to the end of the 70's.  The
80's were equally exciting for me, but this is a topic for a different
conference.

Writing this article  brought to my mind the phrase  of the old German
romanticist, Novalis.  He said: The greatest magician is ``the one who
would cast over himself a spell so complete, that his fantasmagorias would
become autonomous appearances.''

I very much hope that there are many beautiful fantasmagorias ahead  of us.

\endpage

\leftline{\bf References}
\medskip
\item{[1]} L. Landau, A. Abrikosov, and I. Khalatnikov, {\it DAN \bf 95}, 497
(1953).

\item{[2]} M. Gell-Mann and F. Low, {\it Phys. Rev. \bf 111}, 582 (1954).

\item{[3]} A. Anselm, {\it ZHETF \bf 3G}, 863 (1959).

\item{[4]} Y. Nambu and G. Jona-Lasinio, {\it Phys. Rev. \bf 122}, 345
(1961).

\item{[5]} V. Vaks and A. Larkin, {\it ZHETF \bf 40}, 282 (1961).

\item{[6]} A. Migdal and A. Polyakov, {\it ZHETF \bf 51}, 135 (1966).

\item{[7]} V. Gribov and A. Migdal, {\it ZHETF \bf 55}, 1498 (1968).

\item{[8]} A. Polyakov, {\it ZHETF \bf 55}, 1026 (1968).

\item{[9]} A. Polyakov, {\it ZHETF \bf 57}, 271 (1969).

\item{[10]} L. Kadanoff, {\it Phys. Rev. Lett. \bf 23}, 1430 (1969).

\item{[11]} K. Wilson, {\it Phys. Rev. \bf 179}, 1499 (1969).

\item{[12]} A. Polyakov, {\it ZHETF \bf 59}, 542 (1970); {\it ZHETF \bf
60}, 1572 (1971); {\it ZHETF \bf 61}, 2193 (1971).

\item{[13]} A. Polyakov, {\it ZHETF \bf 66}, 23 (1974).

\item{[14]} D. Gross and F. Wilczek, {\it Phys. Rev. Lett. \bf 30}, 1343
(1973).

\item{[15]} H. Politzer, {\it Phys. Rev. Lett. \bf 30}, 1346 (1973).

\item{[16]} D. Amati and M. Testa, {\it Phys. Lett. }

\item{[17]} V. Berezinsky, {\it ZHETF \bf 61}, 1144 (1971).

\item{[18]} A. Polyakov, {\it Phys. Lett. \bf 59B}, 85 (1975).

\item{[19]} A. Polyakov, {\it Phys. Lett. \bf B72}, 477 (1978).

\item{[20]} L. Susskind, {\it Phys. Rev. \bf D20}, 2610 (1979).

\item{[21]} A. Polyakov, {\it Proceedings of Lepton-Photon Symposium},
Batavia (1979).

\item{[22]} A. Polyakov, {\it Nucl. Phys. \bf B164}, 171 (1980).

\item{[23]} Yu Makeenko and A. Migdal, {\it Nucl. Phys. \bf B188}, 269
(1981).

\end